\documentclass[onecolumn, superscriptaddress, preprint]{revtex4-1}
\usepackage{graphicx}

\begin{document}

\title{Evolution of charge order topology across a magnetic phase transition in cuprate superconductors}%

\author{Mingu Kang}%
\affiliation {Department of Physics, Massachusetts Institute of Technology, Cambridge, Massachusetts 02139, USA}
\author{Jonathan Pelliciari}%
\affiliation {Department of Physics, Massachusetts Institute of Technology, Cambridge, Massachusetts 02139, USA}
\author{Alex Frano}%
\affiliation {Department of Physics, University of California, Berkeley, California 94720, USA.}
\affiliation {Department of Physics, University of California, San Diego, California 95203, USA.}
\author{Nicholas Breznay}%
\affiliation {Department of Physics, University of California, Berkeley, California 94720, USA.}
\author{Enrico Schierle}
\affiliation {Helmholtz-Zentrum Berlin für Materialien und Energie, Albert-Einstein Stra$\beta$e 15, D-12489 Berlin, Germany.}
\author{Eugen Weschke}
\affiliation {Helmholtz-Zentrum Berlin für Materialien und Energie, Albert-Einstein Stra$\beta$e 15, D-12489 Berlin, Germany.}
\author{Ronny Sutarto}
\affiliation {Canadian Light Source, Saskatoon, Saskatchewan S7N 2V3, Canada.}
\author{Feizhou He}
\affiliation {Canadian Light Source, Saskatoon, Saskatchewan S7N 2V3, Canada.}
\author{Padraic Shafer}
\affiliation {Advanced Light Source, E. O. Lawrence Berkeley National Laboratory, Berkeley, California 94720, USA.}
\author{Elke Arenholz}
\affiliation {Advanced Light Source, E. O. Lawrence Berkeley National Laboratory, Berkeley, California 94720, USA.}
\author{Mo Chen}
\affiliation {Department of Physics, University of California, Berkeley, California 94720, USA.}
\author{Keto Zhang}
\affiliation {Department of Physics, University of California, Berkeley, California 94720, USA.}
\author{Alejandro Ruiz}
\affiliation {Department of Physics, University of California, Berkeley, California 94720, USA.}
\author{Zeyu Hao}
\affiliation {Department of Physics, University of California, Berkeley, California 94720, USA.}
\author{Sylvia Lewin}
\affiliation {Department of Physics, University of California, Berkeley, California 94720, USA.}
\author{James Analytis}
\affiliation {Department of Physics, University of California, Berkeley, California 94720, USA.}
\author{Yoshiharu Krockenberger}
\affiliation {NTT Basic Research Laboratories, NTT corporation, 3-1 Morinosato-Wakamiya, Atsugi, Kanagawa 243-0198, Japan.}
\author{Hideki Yamamoto}
\affiliation {NTT Basic Research Laboratories, NTT corporation, 3-1 Morinosato-Wakamiya, Atsugi, Kanagawa 243-0198, Japan.}
\author{Tanmoy Das}
\affiliation {Department of Physics, Indian Institute of Science, C. V. Raman Road, Bangalore 560012, India.}
\author{R. Comin}%
\email{rcomin@mit.edu}
\affiliation {Department of Physics, Massachusetts Institute of Technology, Cambridge, Massachusetts 02139, USA}

\begin{abstract}
Charge order is now accepted as an integral constituent of cuprate high-temperature superconductors, one that is intimately related to other instabilities in the phase diagram including antiferromagnetism and superconductivity \cite{1,2,3,4,5,6,7,8,9,10,11}. Unlike nesting-induced Peierls-like density waves, the charge correlations in the CuO$_{2}$ planes have been predicted to display a rich momentum space topology depending on the detailed fermiology of the system \cite{12,13,14,15,16,17,18,19}. However, to date charge order has only been observed along the high-symmetry Cu-O bond directions. Here, using resonant soft X-ray scattering, we investigate the evolution of the \textit{full momentum space topology} of charge correlations in \textit{T}’-Ln$_{2}$CuO$_{4}$ (Ln=Nd, Pr) as a function of intrinsic electron doping. We report that, upon electron doping the parent Mott insulator, charge correlations first emerge in a hitherto-unobserved form, with full (\textit{C}$_{inf}$) rotational symmetry in momentum-space. At higher doping levels, the orientation of charge correlations is sharply locked to the Cu-O bond high-symmetry directions, restoring a more conventional bidirectional charge order with enhanced correlation lengths. Through charge susceptibility calculations, we closely reproduce the drastic evolution in the topology of charge correlations across an antiferromagnetic quantum phase transition, highlighting the interplay between spin and charge degrees of freedom in electron-doped cuprates. Finally, using the established link between charge correlations and the underlying fermiology, we propose a revised phase diagram of \textit{T}’-Ln$_{2}$CuO$_{4}$ with a superconducting region extending toward the Mott limit.
\end{abstract}

\maketitle


In cuprates, an unconventional momentum-space electronic structure rapidly surfaces upon doping carriers into the parent Mott insulator. The unconventional fermiology of lightly hole-doped cuprates is embodied in a strongly momentum-dependent ‘pseudogap’. The pseudogap modifies the single-particle spectral function and Fermi surface, with the latter exhibiting a disconnected topology characterized by four distinctive ‘Fermi arcs’ \cite{20,21,22} In this emergent state, electronic carriers in the CuO$_{2}$ planes concomitantly self-arrange into periodic structures (or charge order) in real space \cite{1,2,3,4,5,6,7,8,9,10,11}. The visualization and elucidation of the microscopic link between these density waves and the many-body fermiology is a stepping stone toward understanding the nature of charge order and its relationship to the Mott physics, pseudogap, and high-temperature superconductivity. In the hole-doped cuprates, charge order has been mapped out to a great extent \cite{11} yet its origin and driving mechanism remain unclear. In Bi-based compounds, scanning tunneling microscopy (STM) \cite{2,3,6,7} and resonant X-ray scattering (RXS) \cite {6,7} experiments proposed that an instability of the pseudogapped Fermi surface might underlie the development of charge ordering. This proposal captures the temperature and doping dependence of charge order \cite{6,7,11}, and has been supported from charge susceptibility calculations in various implementations \cite{17,18,19}. On the other hand, recent STM and resonant inelastic X-ray scattering studies detect charge order outside the pseudogap phase in the very underdoped \textit{p}$\approx$0.07/Cu and overdoped \textit{p}$\approx$0.23/Cu limits (here \textit{p} is the number of holes per Cu), leaving the exact relationship between the charge order and fermiology unsettled \cite{23,24}.

Electron-doped cuprates are an alternative platform to gain new perspectives on this problem, given their analogies to hole-doped cuprates in the phenomenology of charge order \cite{9,10}. At variance with hole-doped systems, here antiferromagnetic (AFM) correlations persist over a wide doping range (Fig. 1a), and deeply influence the many-body fermiology of electron-doped cuprates, as reflected in the appearance of an AFM pseudogap and the evolution of Fermi surface topology with doping \cite{22,25}. Further, in these systems charge order populates a doping region characterized by strong AFM correlations \cite{9,10}, setting an ideal stage to study the interplay between the spin and charge degrees of freedom in cuprates.

The intimate relationship between AFM and charge order has been explored in numerous theoretical studies \cite{12,13,14,15,16} that have focused on density wave instabilities localized along the Cu-O bond directions, in accordance with experimental reports \cite{1,2,3,4,5,6,7,8,9,10,11}. At the same time, theory suggests that strong AFM correlations might induce charge instabilites in broader regions of momentum space (\textbf{Q}-space) \cite{13,14,16}. Charting the topology of charge correlations in electron-doped cuprates is thus a new opportunity to probe the connection between charge order, antiferromagnetism, and many-body fermiology in cuprates.

In this paper, we investigate the evolution of the \textbf{Q}-space topology of charge correlations in \textit{T}’-Ln$_{2}$CuO$_{4}$ (Ln=Nd, Pr) thin films as a function of electron doping. In contrast to widely studied Ce-doped superconductors (\textit{T}’-Ln$_{2-x}$Ce$_{x}$CuO$_{4}$), thin films of \textit{T}’-Ln$_{2}$CuO$_{4}$  can host superconductivity without cation substitution, when subjected to post-growth reduction annealing procedures \cite{26}. The possibility of superconductivity in the ‘undoped’ limit questions the very Mott nature of the parent \textit{T}’-cuprates, and alternative scenarios involving a Slater mechanism have been proposed \cite{26,27}. On the other hand, recent angle-resolved photoemission spectroscopy (ARPES), X-ray photoemission spectroscopy (XPS), and X-ray absorption spectroscopy (XAS) studies concluded that superconducting \textit{T}’-Ln$_{2}$CuO$_{4}$ are actually electron-doped, presumably due to oxygen vacancies \cite{28,29,30}. However, the exact doping-temperature phase diagram remains unclear due to difficulties in estimating the intrinsic electron concentration in these samples.

Figure 1b illustrates the schematics of our experimental setup. In RXS experiments, the in-plane component of momentum transfer (Q$_{||}$) is scanned by rotating the sample about the axis perpendicular to the scattering plane ($\theta$). When the crystallographic \textbf{a} or \textbf{b} axes lie in the scattering plane, scans of the $\theta$ angle probe charge correlations along the Cu-O bond directions, where bidirectional charge order has been so far observed \cite{2,3,4,5,6,7,8,9,10,11}. To access the full momentum (\textbf{Q}) space structure of charge correlations, we performed successive $\theta$ scans for different orientations of the CuO$_{2}$ planes (obtained by varying the azimuthal angle, $\phi$), providing access to the scattering information in the \textbf{Q}-space region between the reciprocal \textbf{H} and \textbf{K} axes. The range of \textbf{Q}-space accessed by this method is schematically shown in the inset of Fig. 1b.

We first demonstrate the presence of enhanced charge correlations in \textit{T}’-Nd$_{2}$CuO$_{4}$ (hereafter referred to as \textit{T}’-NCO\#1) along the Cu-O bond directions. Figures 1c,d show the series of RXS scans as a function of photon energy and temperature. When the photon energy is tuned to the maximum of the Cu-\textit{L}$_{3}$ absorption edge (E $\approx$ 932 eV), the scattered intensity displays clear peaks in momentum space centered at $\arrowvert$Q$\arrowvert$$_{||}$ $\approx$ 0.2 reciprocal lattice units, or r.l.u. (Fig. 1c). The scattering peak rapidly vanishes as the photon energy is tuned off resonance, confirming it originates from electrons in the CuO$_{2}$ planes. As displayed in Fig. 1d, the intensity of this peak smoothly diminishes with increasing temperature, but survives up to room temperature, as in other electron-doped cuprates \cite{9,10}. After subtracting the room temperature background from the 12 K data (inset of Fig. 1d), we extract a correlation length $\xi$ $\approx$ 5-6 unit cells or $\approx$ 20-25 Å. Such short-ranged charge correlations have been seen in other cuprates, (Nd,La)$_{2-x}$Ce$_{x}$CuO$_{4}$ (NCCO, LCCO), Bi$_{2}$Sr$_{2-x}$La$_{x}$CuO$_{6+\delta}$ (BSLCO), and HgBa$_{2}$CuO$_{4+\delta}$ (HgBCO) \cite{7,8,9,10}.

The presence of periodic charge modulations in CuO$_{2}$ planes indirectly reflects the intrinsic carrier doping in the chemically-undoped \textit{T}’-Ln$_{2}$CuO$_{4}$ as previously evidenced from ARPES, XPS, and XAS experiments \cite{28,29,30}. The observed wave vector of charge modulations (Q$_{c}$) can be used to estimate the carrier concentration from experimental measurements of Q$_{c}$ vs. doping level in both hole- and electron-doped cuprates \cite{4,7,9,10,11}. Previous RXS studies reported that Q$_{c}$ in electron-doped cuprates (NCCO, LCCO) increases with higher electron content, similarly to hole-doped cuprates (BSLCO, HgBCO, YBCO). From previous estimates \cite{9,10}, the observed wave vector of \textit{T}’-NCO\#1 (Q$_{c}$ $\approx$ 0.2 r.l.u.) maps onto an electron density \textit{n} $\approx$ 0.07$\pm$0.02 / Cu, placing our sample at the low-doping limit of the phase diagram (see also Fig. 4). This assignment of doping level is consistent with temperature-dependent Hall measurements performed on the same \textit{T}’-Ln$_{2}$CuO$_{4}$ films (see Supplementary Information).

In this low-doping limit, we successively performed RXS scans at various azimuthal angles $\phi$ to map the complete \textbf{Q}-space topology of charge correlations. Surprisingly, as shown in Fig. 2a, broad but clear diffraction peaks are detected along all investigated momentum directions $\phi$= 0, 15, 30, 45, 65, and 90$^{\circ}$ (see Supplementary Fig. 1,2 for the full datasets). To investigate this feature in more detail, we subtract a slowly-varying fluorescence background from each scan and plot the residual resonant peaks in Fig. 2b. The peaks display almost identical intensity, wave vector (Q$_{c}$ $\approx$ 0.2$\pm$0.02 r.l.u.), and correlation length (HWHM $\approx$ 0.07$\pm$0.02 r.l.u.), regardless of the probed momentum direction. Furthermore, the detailed temperature and photon-energy dependence of the scattering peaks at representative azimuthal angles (0, 15, and 45$^{\circ}$, shown in Fig. 2g,h, and Supplementary Fig. 2) fall into a single curve within experimental errors, indicating these structures arise from a common origin. The two-dimensional plot in Fig. 2c summarizes this observation, highlighting the fact that, at low doping levels, the charge correlations in the CuO$_{2}$ plane possess full (\textit{C}$_{inf}$) rotational symmetry in \textbf{Q}-space. This result is in marked contrast to all previous reports of bond-oriented charge order in cuprates (Fig. 2d) \cite{4,5,6,7,8,9,10,11}, and demonstrates that upon electron doping to the parent Mott/Slater insulator, charge correlations initially develop a complex \textbf{Q}-space structure that was entirely unanticipated. To rule out an extrinsic origin to the observed structure factor, the single crystallinity (or \textit{C}$_{4}$ symmetry) of the sample as well as the absence of a diffuse scattering signal have been confirmed before and after the RXS experiments (see Supplementary Fig. 3). Therefore, the charge correlations in \textit{T}’-NCO\#1 manifest a higher rotational symmetry than the underlying lattice.

To gain additional insights on the observed topology of the structure factor S(\textbf{Q}), we calculate the real space correlation function $C(\mathbf{r})=\langle\delta\rho(\mathbf{r}+\mathbf{r_{0}})\delta\rho(\mathbf{r_{0}})\rangle_{\mathbf{r_{0}}}$ and simulate a possible density-wave field $\delta\rho(\textbf{r})$. Note that while $C(\textbf{r})$ is uniquely determined from $S(\textbf{Q})$, $\delta\rho(\textbf{r})$ is generated synthetically from randomly-assigned reciprocal space complex phases (and thus represents only one of the possible realizations; see Methods). As shown in Supplementary Fig. 4, the correlation function $C(\textbf{r})$ exhibits damped oscillations along the radial direction, whose functional form resembles the Bessel function $J_{0}(r)$. The simulated map of the charge density fluctuation $\delta\rho(\textbf{r})$ in Fig. 2e depicts a ‘glassy’ state with an apparent tendency to periodic ordering at wave vector Q$_{c}$ (or 5 unit cells periodicity) but without any orientational preference in contrast to the case of bond-oriented charge order (Fig. 2f).

To elucidate this new phenomenology, we recall that, within Lindhard theory \cite{31}, the linearized charge density perturbation $\rho(\mathbf{Q})$ induced by an external potential $\phi(\mathbf{Q})$ is obtained as $\rho(\mathbf{Q})=\chi(\mathbf{Q})\cdot\phi(\mathbf{Q})$, where the Lindhard function or generalized charge susceptibility $\chi(\mathbf{Q})$ is directly linked to the underlying electronic band structure. In this framework, a local potential can induce Friedel oscillations akin to quasi-particle interference modulations observed in STM, and RXS peaks can emerge even in the absence of genuine long-range ordering, as revealed by recent theoretical studies \cite{18,32,33}. Given that the experimental observable, the structure factor $S(\mathbf{Q})$, is proportional to the squared amplitude of the charge density (i.e. $\arrowvert\rho(\mathbf{Q})\arrowvert^{2}=\arrowvert\chi(\mathbf{Q})\cdot\phi(\mathbf{Q})\arrowvert^{2}$), this formulation of the scattering process allows to analyze the RXS information not only in terms of spontaneous symmetry breaking, but also as a direct reflection of generic many-body instabilities in Fermi surface, encoded in the interacting charge susceptibility.

To articulate this scenario, in Fig. 3 we compare the Fermi surface topology, the \textbf{Q}-space maps of the static charge susceptibility, and the \textbf{Q}-space topology of observed RXS peaks. Figures 3a-e and Figures 3f-j respectively show the calculated Fermi surfaces and static charge susceptibilities of \textit{T}’-NCO at various representative doping levels. The calculation is based on the momentum-resolved density fluctuation (MRDF) model, where the spin and charge density correlation spectrum is computed self-consistently near the AFM saddle-point with momentum-dependent self-energy correction \cite{34}. This model has reproduced certain experimental characteristics of charge order in hole-doped cuprates from the instabilites in the Fermi surface \cite{19}. In electron-doped systems, the longitudinal spin and charge susceptibilities become mixed in the presence of commensurate AFM correlation \cite{34,35}, which results in the spin-charge interaction being distinct from the incommensurate stripe physics of hole-doped cuprates (see Supplementary Information for details). In the low-doping limit (\textit{n} = 0.05, Fig. 3a) with strong AFM correlations, the electrons first fill the upper Hubbard states at antinodes, developing electron pockets centered at ($\pm\pi$, 0)/(0, $\pm\pi$). Notably, the electron pockets are very circular in this limit due to the presence of strong ($\pm\pi$, $\pm\pi$) AFM scattering. Correspondingly, the interacting charge susceptibility is enhanced uniformly along all azimuthal directions at $\arrowvert$Q$\arrowvert_{||}$ = 0.2 r.l.u. (Fig. 3f,l), in remarkable accordance with the observed \textbf{Q}-space structure of the scattering peaks in \textit{T}’-NCO\#1 (Fig. 3k,m). This analysis demonstrates that the RXS structures in Fig. 2 reflect the rotationally symmetric particle-hole scattering channels connecting low-energy states across the Fermi surface, which materialize in real space possibly through the defect-induced Friedel oscillations in charge density \cite{18,32,33}.

The dramatic change in Fermi surface topology occurs slightly below the AFM quantum critical point, where the termination of Néel order accompanies the collapse of the Mott gap, resulting in the formation of hole pockets at ($\pm\pi$/2, $\pm\pi$/2) (see Fig. 3c,d). With the weakening of ($\pm\pi$, $\pm\pi$) AFM scattering, the charge susceptibility becomes anisotropic, and is enhanced around the high-symmetry Cu-O bond directions (Fig. 3h,i), recovering the \textit{C}$_{4}$ symmetry of the underlying lattice. This trend continues up to the paramagnetic limit (\textit{n} = 0.17, Fig. 3e), where the electron and hole pockets merge to form a single large Fermi surface centered at ($\pm\pi$, $\pm\pi$). We note that the doping evolution of the Fermi surface presented here is in close agreement with experimental Fermi surfaces obtained from ARPES and quantum oscillation experiments on electron-doped cuprates \cite{22,25}. These calculations thus suggest that AFM order in electron-doped cuprates promotes charge correlations in all momentum directions in the low-doping limit as detected by the present RXS experiments, while a more conventional bidirectional charge order emerges as AFM order is removed at larger doping.

To confirm the crossover from \textit{C}$_{inf}$ to \textit{C}$_{4}$ topology of the charge correlation, we performed RXS measurements on the \textit{T}’-Pr$_{2}$CuO$_{4}$ (\textit{T}’-PCO\#1) sample with higher intrinsic doping. As displayed in Figure 3o, we observed the sharp peak in RXS intensity at $\arrowvert$Q$\arrowvert_{||}$ = 0.32$\pm$0.01 r.l.u. along the Cu-O bond direction. From the empirical doping dependence of Q$_{c}$ – or equivalently by comparing Q$_{c}$ to the maximum wave vector of the calculated charge susceptibility (Fig. 3n) – Q$_{c}$ = 0.32$\pm$0.02 r.l.u. maps to an electron density n $\approx$ 0.17$\pm$0.02 \cite{10} (See also Supplementary Information for the corresponding Hall measurements). In this large-doping limit, the RXS peak entirely disappears when measured slightly off from the Cu-O bond axis (Fig. 3o) indicating that the charge correlations are confined to the ($\pm\pi$,0)/(0,$\pm\pi$) directions (See the Supplementary Fig. 5 for details of RXS experiments on \textit{T}’-PCO\#1). This observation confirms that low-doping charge correlations with \textit{C}$_{inf}$ rotational symmetry morph into a disconnected topology with charge order peaks pinned to the Cu-O bond directions in the high-doping limit, forming a more conventional type of bidirectional charge order as schematically displayed in Fig. 3p. 

It is worth noting that the charge order in the high-doping limit (\textit{T}’-PCO\#1) is rather long-ranged, with a correlation length $\xi$ $\approx$ 20 unit cells or $\approx$ 80 Å, which is ~4 times larger than the low-doping limit (\textit{T}’-NCO\#1). This length scale is comparable to that in YBa$_{2}$Cu$_{3}$O$_{6+d}$ (YBCO), where charge order directly competes with superconductivity \cite{4}. This fact hints at a possible crossover in the nature of charge order as a function of doping: from an incipient ordering phenomenon in the low-doping regime, to a long-range-ordered ground state at higher electron doping. Our observation is also consistent with the proposals that the length scale of charge order is governed by the degree of directionality in the underlying electronic susceptibility \cite{18}.

On the other hand, the short correlation length of \textit{T}’-NCO\#1 is similar to that of BSLCO and HgBCO \cite{7,8}, where the direct competition between charge order and superconductivity is not fully established. This analogy raises an important question on the nature of charge order in these hole-doped systems: whether it is a \textit{bona fide} symmetry-broken phase or rather an incipient state possibly initiated by quasiparticle scattering across the Fermi surface as observed here. At the same time, charge correlations in hole-doped cuprates are not expected to exhibit a complex \textbf{Q}-space topology owing to the less prominent AFM correlations and a more pronounced particle-hole asymmetry in the band structure. Correspondingly, STM experiments on BSLCO have recently found checkerboard order down to a hole doping \textit{p} $\approx$ 0.07, suggesting that the charge correlations in hole-doped cuprates retain a bidirectional form at all doping levels \cite{23}. Altogether, our results combined with previous reports on hole-doped cuprates \cite{4,5,6,7,8,11}, hint at a common link between the low-energy fermiology and the spatial organization of the electronic state.

The established connection paves the way to use Q$_{c}$ as a measure of intrinsic electron density. This inference is especially valuable for all \textit{T}’-structured cuprates, whose unknown oxygen stoichiometry usually hinders the exact determination of their true electron density \cite{26,28,29,30,36,37}. In Fig. 4b, we plot the calculated doping dependence of Q$_{c}$ as extracted from the maximum of the charge susceptibility (Fig. 4a). For comparison, we also plot the Ce-doping dependence of Q$_{c}$ in NCCO and LCCO from previous studies \cite{10}. A general agreement in trend is obtained, nonetheless the NCCO data appear to be slightly offset to higher doping levels. To some degree, this deviation is natural since the Ce concentration cannot represent the sole factor determining carrier density in electron-doped cuprates, as also emphasized in recent photoemission studies \cite{28,29,30,36,37}. This very fact further underscores the importance of the explicit determination of intrinsic electron concentrations in \textit{T}’-cuprates. The established relationship between Q$_{c}$ and \textit{n} introduces a possible method to gauge doping levels in these systems and is applicable to both bulk and thin films.

In addition to \textit{T}’-NCO\#1 (Q$_{c}$ $\approx$ 0.2 r.l.u.) and \textit{T}’-PCO\#1 (Q$_{c}$ $\approx$ 0.32 r.l.u.) presented above, we measured two additional samples with RXS peaks at Q$_{c}$ $\approx$ 0.22 r.l.u. (\textit{T}’-PCO\#2) and Q$_{c}$ $\approx$ 0.25 r.l.u. (\textit{T}’-NCO\#2) (Supplementary Fig. 6). Based on the relationship between Q$_{c}$ and \textit{n}, we infer that our samples cover a wide doping range from \textit{n} $\approx$ 0.05 to 0.17 as marked in Fig. 4b. Interestingly, all our samples were subjected to the post-growth annealing and show superconductivity with T$_{C}$ = 24.4, 23.1, 24.6, and 24.8 K (for \textit{T}’-NCO\#1, \textit{T}’-PCO\#2, \textit{T}’-NCO\#2, and \textit{T}’-PCO\#1, respectively). As shown in Fig. 4c, this finding indicates a persistence of the superconducting phase into the low-doping region, in contrast to the conventional phase diagram of electron-doped cuprates (Fig. 1a). This conclusion lends support to the very recent ARPES studies on bulk \textit{T}’-Pr$_{2-x}$LaCe$_{x}$CuO$_{4-d}$, which also reports a revised phase diagram with an extended superconducting dome based on the intrinsic electron density estimated from Luttinger’s theorem \cite{36,37}.

In sum, we have observed the crossover in the topology of charge correlations in cuprates as a function of electron doping across the AFM quantum critical point. In doing so, we have uncovered a direct link between the fermiology and the empirical charge correlations in electron-doped cuprates. Based on the intrinsic electron density estimated from the characteristic wave vector of the charge modulations, we have reconstructed the doping-temperature phase diagram of \textit{T}’-Ln$_{2}$CuO$_{4}$ and revealed an extended superconducting phase toward the underdoped region. Altogether, these experimental observations challenge the conventional understanding of charge order in cuprates, and provide new crucial insights onto a unified description of this phenomenology.

\newpage

\section*{Methods}
\textbf{Sample growth and characterizations} Thin films of \textit{T}’-NCO and \textit{T}’-PCO used in this work were grown by molecular beam epitaxy under ultra-high vacuum using Nd, Pr, and Cu metal sources and atomic oxygen generated in-situ from an RF oxygen source \cite{38}. Reflection high energy electron diffraction (RHEED) and electron impact emission spectroscopy (EIES) were used to monitor and control the growth of NCO and PCO films on (001) SrTiO$_{3}$ substrates in real time. High-resolution reciprocal space mapping data show that the films of thickness $\approx$ 100 nm are grown fully relaxed. The films were subjected to the two-step annealing process \cite{39}. The superconducting transition was measured by electrical transport and magnetometry.

\textbf{Resonant X-ray scattering experiments} RXS experiments were performed at the UE46 PGM-1 beamline of the BESSY II (\textit{T}’-NCO\#1, \textit{T}’-NCO\#2), the REIXS beamline of the Canadian Light Source (\textit{T}’-PCO\#1), and the Beamline 4.0.2 of the Advanced Light Source (\textit{T}’-PCO\#2). Experiments were conducted at 12 K (unless specified) and under a high vacuum better than 10$^{-9}$ Torr. Samples were oriented \textit{in situ} using Bragg reflections. All measurements were conducted at the Cu-\textit{L}$_{3}$ absorption edge with out-of-scattering-plane ($\sigma$) incoming polarization to maximize the sensitivity to charge scattering. Momentum-space scans were obtained by rocking the sample angle at fixed detector position. The temperature dependence series were acquired by both heating and cooling the samples, yielding consistent results.

\textbf{Simulations of the real space charge fluctuations} The structure factor \textit{S}(\textbf{Q}) for \textit{T}’-NCO\#1 is modeled by \textit{S}(\textbf{Q})=exp(4 $ln$2 ($\arrowvert$Q$\arrowvert$-Q$_{c}$)$^{2}$/$w^{2}$), where the Q$_{c}$ is the wave vector centroid of charge correlations, and $w$ is the FWHM. \textit{S}(\textbf{Q}) for BSLCO is similarly modeled by a 2D Gaussian function centered at ($\pm$Q$_{c}$, 0), (0, $\pm$Q$_{c}$). The real space charge-charge correlation function \textit{C}(\textbf{r}) is obtained by discrete 2D Fourier transform of \textit{S}(\textbf{Q}). To simulate the real space charge density fluctuation $\delta\rho(\textbf{r})$, we first introduce a matrix of random reciprocal-space phases $\varphi$(\textbf{Q}$_{m}$) and subsequently simulate the charge fluctuation as $\delta\rho(\textbf{r})=\sum_{m}\sqrt{\textit{S}(\textbf{Q}_{m})} cos(\textbf{Q}_{m}\cdot\textbf{r}+\varphi(\textbf{Q}_{m}))$, where \textit{m} labels the discretized Fourier components \cite{40}.

\textbf{Momentum-resolved Density Fluctuation (MRDF) calculations} We take an effective Hamiltonian near the AFM saddle point

\begin{equation}
H=\sum_{k,\sigma}\lbrack\xi_{k}c^{\dag}_{k,\sigma}c_{k,\sigma} + Um\sigma c^{\dag}_{k+Q,\sigma}c_{k,\sigma}\rbrack +h.c.
\end{equation}

where the non-interacting dispersion $\xi_{k}$ is taken from tight-binding calculation of previous works, $U$ is on-site Hubbard interaction, and $m$ gives the staggered magnetization \cite{34}. In the commensurate AFM state with Q=($\pi$,$\pi$), the longitudinal spin and charge susceptibilities become mixed in the Umklapp scattering channels, while the transverse spin susceptibilities remain decoupled. Based on the above Hamiltonian, we compute the correlation functions with many-body corrections implemented within the random-phase approximations and self-energy corrections. In the framework of the MRDF model, both the single-particle Green’s function and the two-body correlation functions are calculated self-consistently with the self-energy corrections. We included Bethe-Salpeter-type vertex corrections obeying the Ward’s identity. The AFM gap and chemical potential are also computed self-consistently for each doping with a doping dependent U adopted from previous studies \cite{34}. See Supplementary Information for a detailed description of the calculation method.

\newpage

\section*{Acknowledgement}
The authors are grateful to S. Kivelson, B. Fine, K.M. Shen, and D. Hawthorn for insightful discussions. The authors thank L. Ye for support with resistivity measurements. This material is based on work supported by the National Science Foundation under Grant No. 1751739. The authors acknowledge the Berlin Electron Storage Ring (BESSY II), the Canadian Light Source and the Advanced Light Source for provision of synchrotron radiation beamtime. Research performed in the Canadian Light Source is funded by the Canada Foundation Innovation, the Natural Sciences and Engineering Research Council of Canada, the University of Saskatchewan, the Government of Saskatchewan, Western Economic Diversification Canada, the National Research Council Canada and the Canada Institutes of Health Research. This research used resources of the Advanced Light Source, which is a DOE Office of Science User Facility under contact no. DE-AC02-05CH11231. M.K. acknowledges a Samsung Scholarship from the Samsung Foundation of Culture. J.P. is financially supported by the Swiss National Science Foundation Early Postdoc Mobility fellowship project no. P2FRP2-171824 and Postdoc Mobility fellowship project no. P400P2-180744. Work by N.B. and J.A. is supported by the Gordon and Betty Moore Foundation’s EPiQS Initiative through Grant GBMF4374. T.D. acknowledges financial support from the Infosys Science foundation under Young investigator Award.

\section*{Author contributions}
M.K., J.P., E.S., A.F., and N.B. conducted the RXS experiments and analyzed the data with help from M.C., K.Z., A.R., Z.H., S.L., and J.A.. E.W., R.S., F.H, P.S., and E.A. maintained the X-ray beamlines and supported RXS experiments. T.D. performed the calculations. Y.K. and H.Y. grew the thin films, performed the transport measurements, and analyzed the data. R.C. conceived the experiment and directed the project. M.K. and R.C. wrote the manuscript with input from all other co-authors.

\section*{Competing financial interests}
The authors declare no competing financial interests.

\section*{Data availability}
The data that support the plots within this paper and other findings of this study are available from the corresponding author upon reasonable request.


\newpage
\begin{figure}
\includegraphics[width =  \columnwidth]{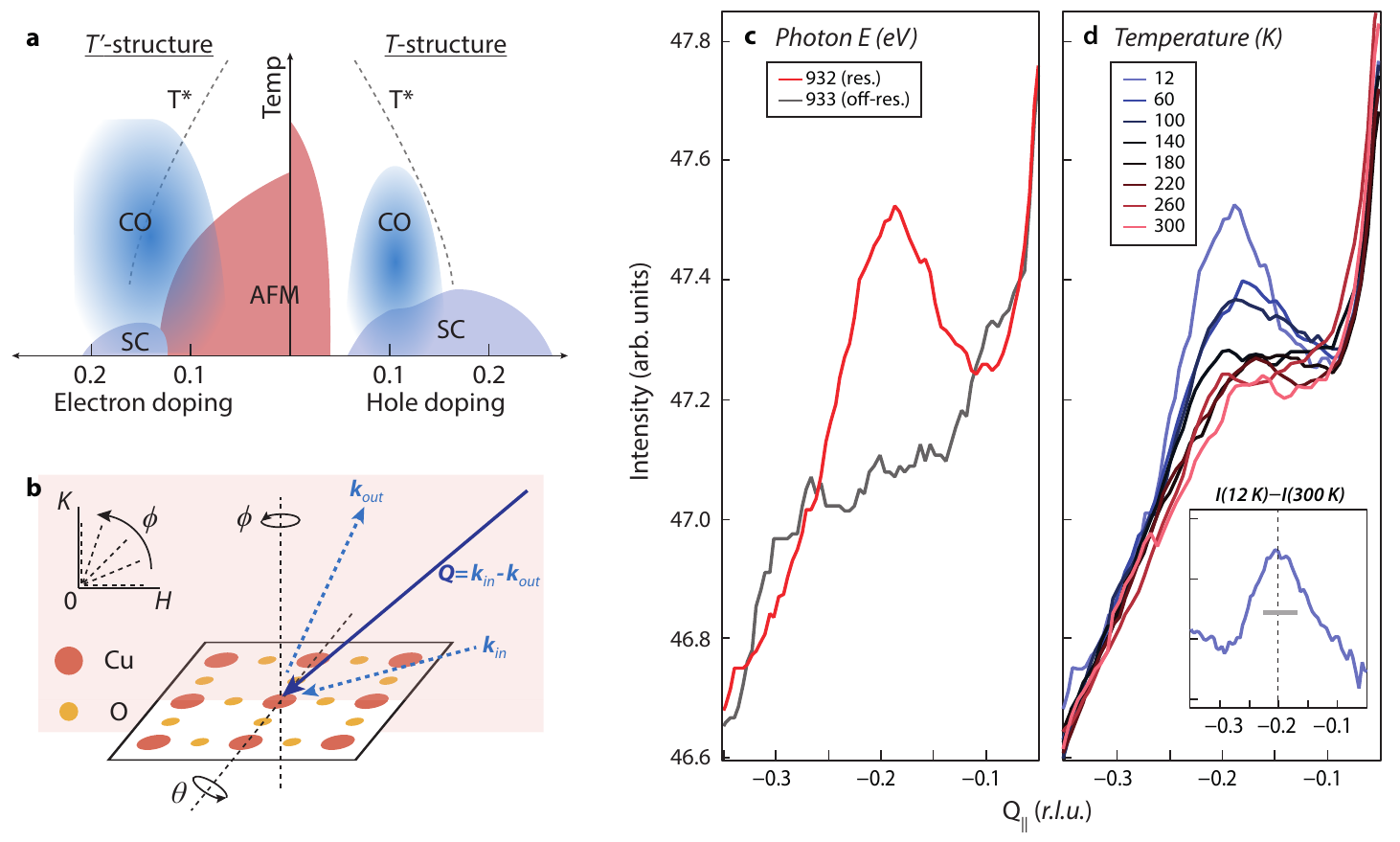}
\caption{\label{fig1} Charge correlations in \textit{T}'-NCO\#1 along the Cu-O bond directions. a, Doping-temperature phase diagram of cuprates with the superconducting (SC), antiferromagnetic (AFM), charge ordering (CO) and pseudogap (T*) regions highlighted. b, Schematics of the RXS experimental geometry. The inset shows the region of \textbf{Q}-space that can be spanned by a combination of in-plane ($\theta$) and out-of-plane ($\phi$) rotations. c,d, Photon energy and temperature dependence of RXS scans along the Cu-O bond direction. Data in c (d) were obtained at 12 K (932 eV). The inset in d shows the 12 K RXS scan after subtracting the room-temperature (300 K) profile. The grey line represents the half-width-at-half-maximum (HWHM).}
\end{figure}

\newpage
\begin{figure}
\includegraphics[width =  \columnwidth]{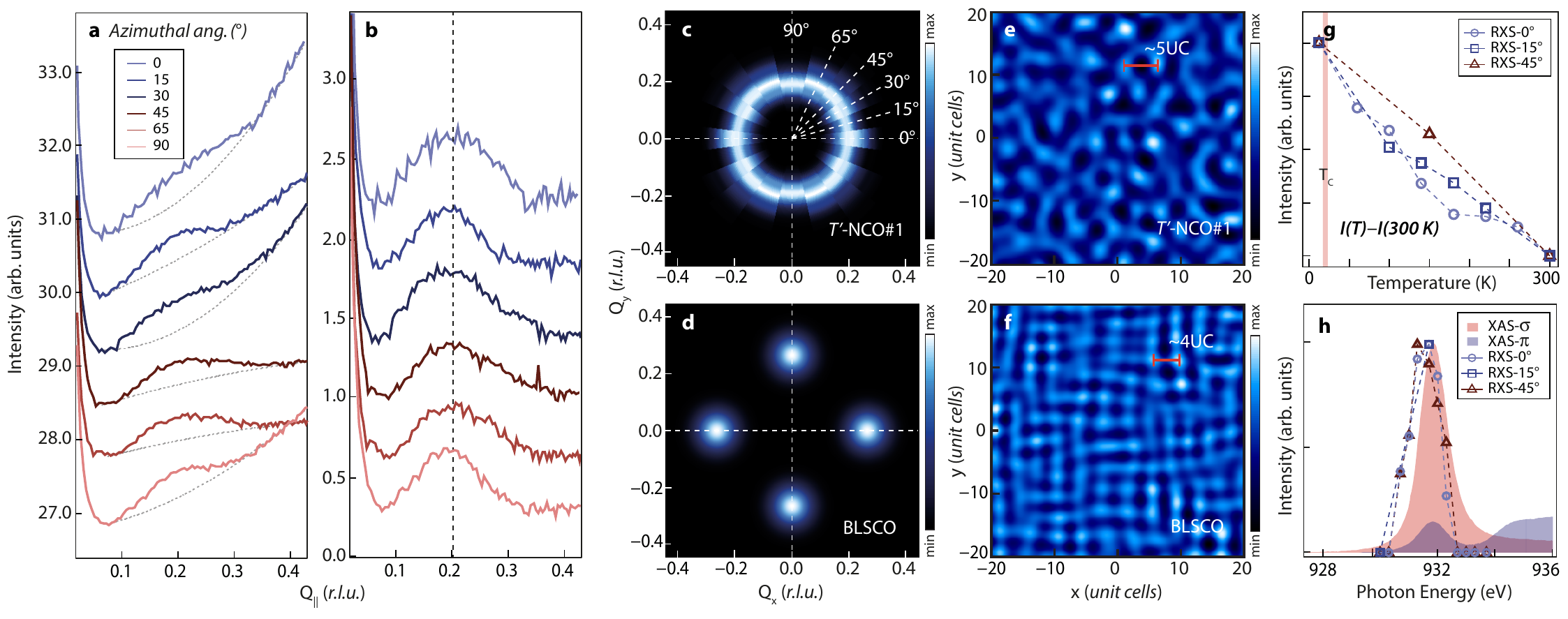}
\caption{\label{fig1} Full momentum-space topology of charge correlations in \textit{T}’-NCO\#1. a, RXS scans of \textit{T}’-NCO\#1 for various azimuthal angles. Curves are vertically shifted for clarity. Overlaid grey dotted-lines represent polynomial fit to the fluorescence backgrounds. b, Charge correlation peaks derived from a after background subtraction. c,d, Momentum space topology of charge correlations (structure factor) in \textit{T}’-NCO\#1 and BSLCO. c is a polar plot of data from b, while d is reproduced from data in Ref.7. e,f, Simulated real space charge density modulations $\delta\rho$(\textbf{r}) corresponding to the structure factors in c and d, respectively. g,h, Temperature and photon energy dependence of RXS intensities at representative azimuthal angles 0, 15, and 45$^{\circ}$. Shaded profiles in h represent the Cu-\textit{L}$_{3}$ edge X-ray absorption scans with $\sigma/\pi$ incoming polarizations.}
\end{figure}

\newpage
\begin{figure}
\includegraphics[width =  \columnwidth]{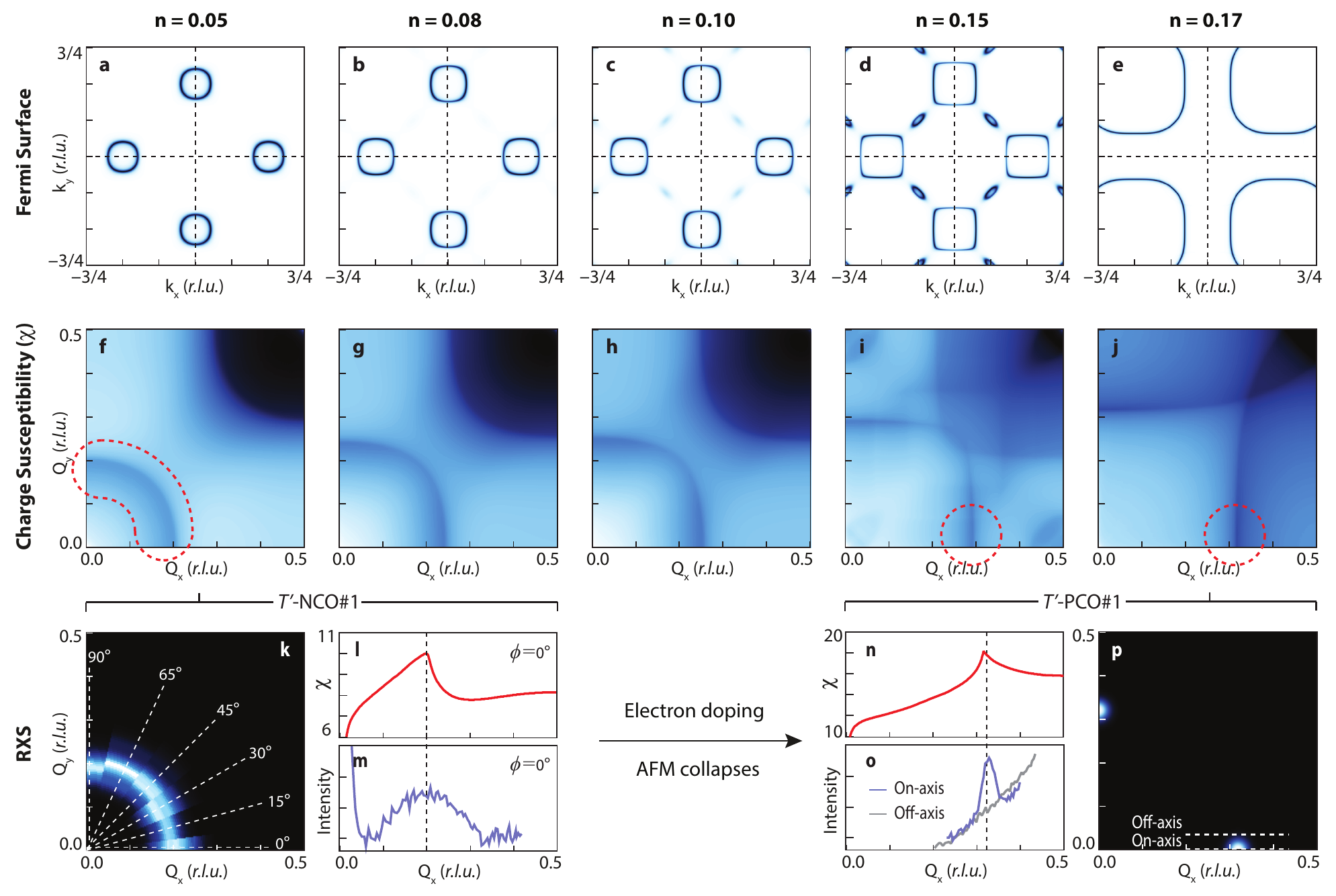}
\caption{\label{fig1} Crossover from \textit{C}$_{inf}$ to \textit{C}$_{4}$ symmetric charge correlation in electron-doped cuprates. a-e, Evolution of the Fermi surface of electron-doped cuprates as a function of intrinsic electron density. f-j, Corresponding \textbf{Q}-space plots of the static charge susceptibilities obtained from MRDF calculations. Red-dotted lines highlight the contours where an enhanced charge response is expected. k,p, Q-space representation of charge correlations in the low-doping (\textit{T}’-NCO\#1, \textit{n} $\approx$ 0.05) and high-doping limit (\textit{T}’-PCO\#1, \textit{n} $\approx$ 0.17). l-o, Comparison of charge susceptibilities (l,n) and RXS intensities (m,o) along Q$_{x}$ directions at \textit{n} $\approx$ 0.05 (l,m) and \textit{n} $\approx$ 0.17 (n,o). Dotted-lines in p represent the momentum-space segments along which the RXS profiles of \textit{T}’-PCO\#1 in o are sampled.}
\end{figure}

\newpage
\begin{figure}
\includegraphics[width =  \columnwidth]{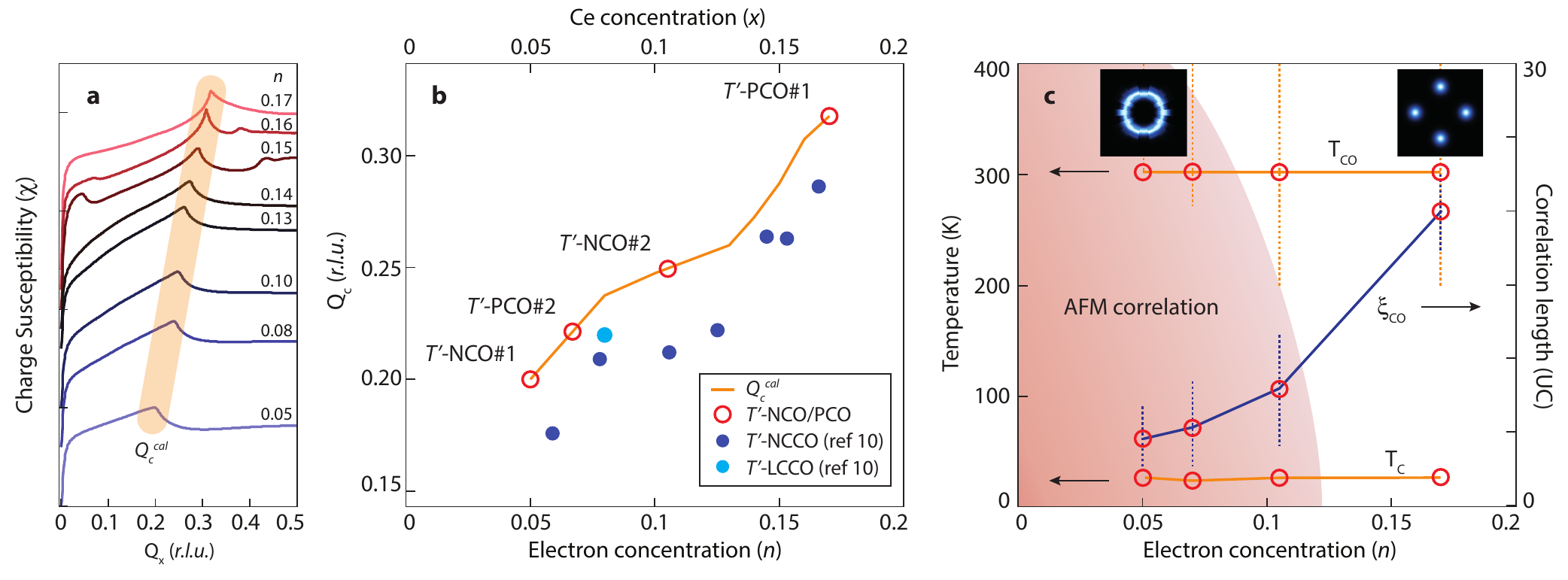}
\caption{\label{fig1} Phase diagram of \textit{T}’-Ln$_{2}$CuO$_{4}$. a, Stack of the charge susceptibility $\chi$ along the reciprocal H-axis calculated at various electron concentration \textit{n}. $\chi$ displays a clear maximum at the wave vector Q$^{cal}_{c}$, which systematically increases with doping. b, Relationship between Q$_{c}$ and dopant concentration \textit{n} in electron-doped cuprates. The orange solid line tracks the Q$^{cal}_{c}$ versus \textit{n}. Dark (light) blue dots represent the Ce-concentration dependence of Q$_{c}$ in NCCO (LCCO) \cite{10}. The four \textit{T}’-Ln$_{2}$CuO$_{4}$ samples investigated here are marked by red circles. c, Temperature-doping phase diagram of \textit{T}’-Ln$_{2}$CuO$_{4}$ with onset temperature of charge correlations (T$_{CDW}$), superconducting transition temperature (T$_{c}$), and AFM region. The latter is sketched on the basis of both neutron scattering \cite{41} and ARPES studies \cite{37}. Orange dotted lines represent error bars on T$_{CDW}$. The insets display the momentum space topology of charge correlations in the low- and high-doping limits. The evolution of $\xi_{CDW}$ as a function of the electron density is also overlaid as a blue solid line.}
\end{figure}

\end{document}